\documentclass{aa}  

\usepackage{graphicx}

\usepackage{txfonts}
\usepackage[colorlinks=true,citecolor=blue,linkcolor=blue]{hyperref}%
\newcommand{\dg}{^{\circ}}
\begin{document}

   \title{Local alignments of parsec-scale AGN radiojets}

   \author{N. Mandarakas
          \inst{1,2}
          \and
          D. Blinov\inst{1,2,3} \and C. Casadio \inst{1,4} \and V. Pelgrims \inst{1,2} \and S. Kiehlmann\inst{1,2} \and V. Pavlidou\inst{1,2} \and K. Tassis\inst{1,2}}

   \institute{Institute of Astrophysics, Foundation for Research and Technology - Hellas, Voutes, 70013 Heraklion, Greece\\
              \email{nmandarakas@physics.uoc.gr}
         \and
             Department of Physics, University of Crete, 70013, Heraklion, Greece\
        \and
            St. Petersburg State University, Universitetsky pr. 28, Petrodvoretz, 198504 St. Petersburg, Russia\
        \and
           Max-Planck-Institut f\"ur Radioastronomie, Auf dem   H\"ugel, 69, D-53121 Bonn, Germany
           }


 
  \abstract
   {Coherence in the characteristics of neighboring sources in 2D and 3D space may suggest the existence of large-scale cosmic structures, which are useful for cosmological studies. Numerous works have been conducted to detect such features in global scales as well as in confined areas of the sky. However, results are often contradictory and their interpretation remains controversial.}
   {We investigate the potential alignment of parsec-scale radio jets in localized regions of the coordinates-redshift space.}
   {We use data from the Astrogeo VLBI FITS image database to deduce jet directions of radio sources. We perform the search for statistical alignments between nearby sources and explore the impact of instrumental biases.}
   {We unveil four regions for which the alignment between jet directions deviates from randomness at a significance level of more than $5\,\sigma$ and is unlikely due to instrumental systematics. Intriguingly, their locations coincide with other known large-scale cosmic structures and/or regions of alignments.}
   {If the alignments found are the result of physical processes, the discovered regions may designate some of the largest structures known to date.}

   \keywords{}

   \maketitle
%

\section{Introduction}\label{sec:intro}

The detection of large cosmic structures has always been of great interest for the field of cosmology. Studying the properties of such structures allows us to test cosmological models and place stringent constraints in their parameters.
The existence of large-scale structures is predicted by the standard $\rm\Lambda CDM$ model and has been verified by cosmological simulations \citep[e.g.][]{Springel2005}. The homogeneity principle of the $\rm\Lambda CDM$ model sets an upper limit to the largest dimension of such structures at $\sim 260 \; h^{-1} Mpc$ \citep{Yadav2010}.
However, observational works have presented discoveries of super-clusters of galaxies, large quasar groups (LQGs) and giant structures highlighted by a distribution of GRBs \citep[e.g.][]{Clowes2013,Balazs2015,Horvath2015}, which seem to violate this principle and pose a challenge to the $\rm\Lambda CDM$ model. On the other hand, there are works which debate this violation \citep[e.g.][]{Nadathur2013} or even question the existence of these structures \citep[e.g.][]{Ukwatta2016,Christian2020}. The discussion is still inconclusive but future missions are expected to give decisive results \citep{Horvath2020}.

Identification of large-scale structures can be achieved by the detection of objects clustering \citep[e.g.][]{Geller1989,Einasto2001} or by studying the coherence of certain characteristics of distant objects. In several occasions, observations of such features have revealed unexpected diversion from uniformity, which could potentially be indicative of the existence of a large-scale cosmic structure dictating the behavior of the sources.
For instance, \cite{Hutsemekers1998,Hutsemekers2001,Hutsemekers2005} revealed coherent orientation of optical polarization vectors of active galactic nuclei (AGN) in certain areas of the sky. Recently, \cite{Pelgrims2019} showed that such alignments are unlikely to occur due to contamination by Galactic dust. Similar works were conducted in the radio regime.
\cite{Joshi2007} studied radio polarization angles of 4290 sources with polarized flux density $\leq$ 1 mJy from the JVAS/CLASS survey \citep{Jackson2007} in large angular scales without finding any significant alignment. \cite{Tiwari2013} studied the same sample, both in large scales and smaller distances. For the same angular scales probed by \cite{Joshi2007} they confirm the lack of alignments, however they detect a strong alignment signal for a distance scale in the order of 150 Mpc. \cite{Pelgrims2015} additionally used  redshift information for the same sample to test the hypothesis of randomness in 2D and 3D and revealed significant deviation from uniformity for three distinct regions, mainly in the 2D case. Two out of three are consistent with regions detected by \cite{Hutsemekers1998,Hutsemekers2001,Hutsemekers2005}.

A possible correlation between polarization vectors of quasars and their environments was investigated by \cite{Hutsemekers2014}. They compared linear optical polarization vectors of quasars with the projected direction of the quasar group to which they belong and found that polarization vectors are either parallel or perpendicular to that direction. Similar trend is observed for radio polarization \citep{Pelgrims2016}.

Several studies were also conducted to search for alignments of radio structures of neighbouring objects, mainly in the 2D sky coordinates space. \cite{Joshi2007} analyzed a sample of 1565 sources and determined the position angles of their jets. No significant alignments were detected. \cite{Taylor2016} identified a sample of 65 sources with radio-jets in the ELAIS N1 field and studied their projected directions. Their results showed a correlation of the jet directions up to an angular separation scales of $1.8\dg$ (53 Mpc at redshift of 1). \cite{Contigiani2017} tested the preferential direction of radio sources for two different samples. For a sample of 30059 sources from the FIRST catalogue \citep{Becker1995} they found $>3\,\sigma$ significant alignment on angular scales of $1.5\dg-2\dg$, which corresponds to 19-39 Mpc for the mean redshift in their sample. However, similar behaviour was not observed in a sample of 11674 extended sources from the TIFR GMRT Sky Survey \citep{Intema2017}. The lack of alignments was explained by the sparsity of this sample. Lately, \cite{Panwar2020} used information from the FIRST catalogue to study possible alignments of sources with small angular separation. They searched for systematic alignments of the axes of radio galaxies at different angular scales.
They found a significant effect for angular scales less than $1\dg$ without imposing any redshift limitation. Morevover, they conducted a 3D analysis for a subsample of sources with known redshift, and found a weak signal ($\sim2\,\sigma$ significance) for $z\le0.5$ in a comoving distance scale of $\sim640$ Mpc.
Recently, \cite{Osinga2020} compiled a sample of 7555 double-lobed radio sources using LoTSS \citep{Shimwell2019} and performed analysis tests in 2D and 3D. They reported significant deviation from uniformity in the 2D analysis at angular scales of about $4\dg$, whereas the 3D analysis did not reveal a similar trend.

In a previous work \citep{Blinov2020} we investigated for the first time the potential global alignments of parsec-scale AGN radio jets in a large portion of the sky. We inspected 14078 sources and identified jet directions for 6388 of them. We determined their position angles and found no significant alignments at scales between $\sim2\dg-15\dg$ which corresponds to a linear scale greater than 60.5 Mpc for the median of our redshift distribution. In this work we use a slightly augmented sample to explore alignments in localized areas of the sky. Position angles are measured North-to-East.

This paper is organized as follows. In Sect.~\ref{sec:data} we describe the sample and the methods used for the jet detection. In Sect.~\ref{sec:align} we introduce the necessary mathematical tools and we describe how the search for alignments was performed. In Sect.~\ref{sec:results} we present our results, discuss the effect of possible biases in our dataset, and compare our findings with regions of cosmological interest. We conclude in Sect.~\ref{sec:conclusions}.

\section{Data sample and reduction}\label{sec:data}

We collected images of compact radio sources from the Astrogeo VLBI FITS image database\footnote{\url{http://astrogeo.org/vlbi_images/}}, which are gathered from various VLBI projects, but mostly from geodetic programs \citep{Petrov2021}. We used the database version of June 10th, 2020. The database contained 15235 sources and a total of 103483 images at various frequency bands. Redshift information used in this work was retrieved from the OCARS catalog \citep{Malkin2018}.

The methods used to determine the jet position angles (PAs) operate in the image plane. Since original VLBI images are convolved with elliptical beams, elongated restoring beams could introduce biases in the jet position angle determination  \citep{Pushkarev2017}. Therefore, we created new images convolved with a median (where multiple epochs are available) circular beam. Given the dimensions of the elliptical beam, $\alpha$ and $\beta$, the radius of the circular beam was calculated as $\sqrt{\alpha\beta}$. In the case where multiple epochs were present, we aligned the images such that the maximum pixel was at the center, and median-stacked them.

\begin{figure*}
   \centering
   \begin{tabular}{ccc}
   \includegraphics[trim={0.0cm 0.0cm 0.3cm 0.0cm},clip,width=0.31\textwidth]{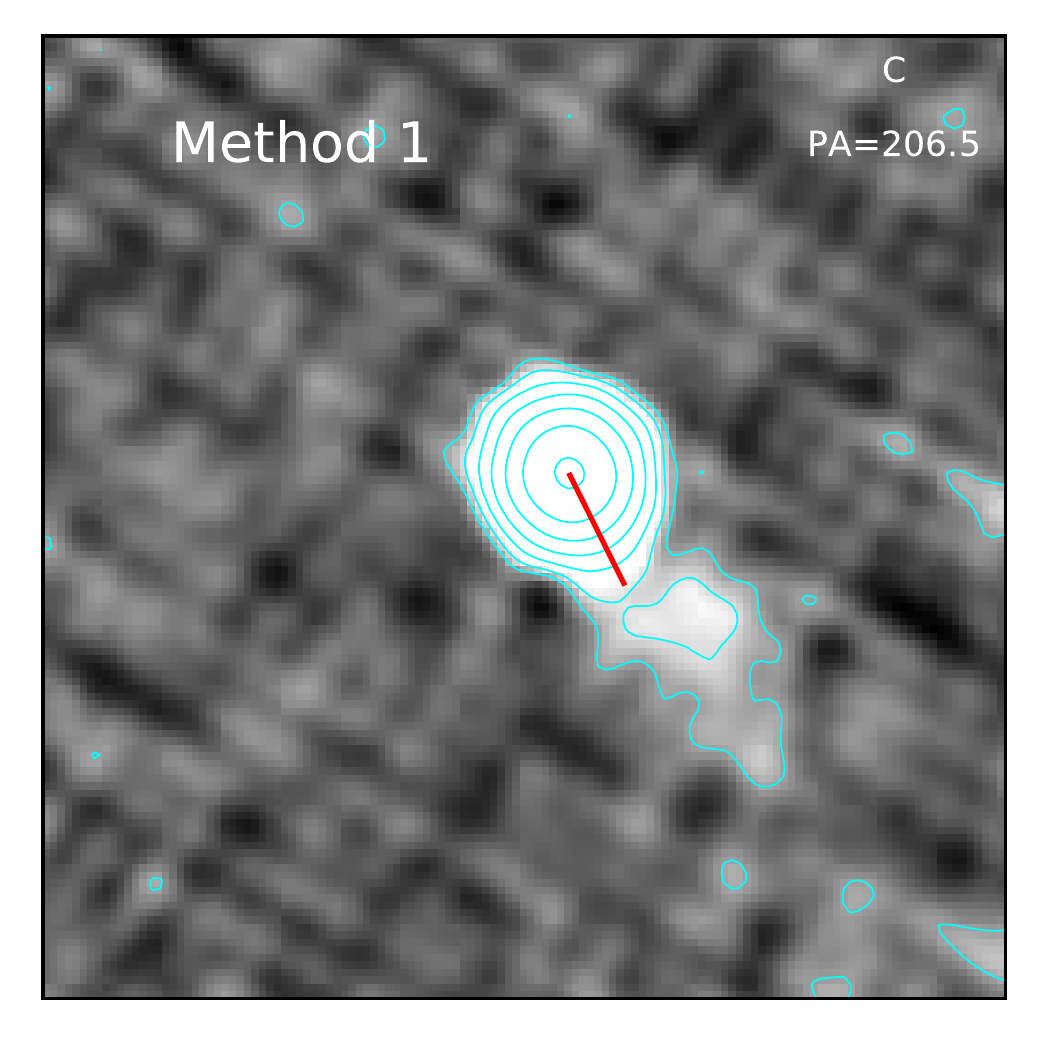}&
   \includegraphics[trim={0.0cm 0.0cm 0.3cm 0.0cm},clip,width=0.31\textwidth]{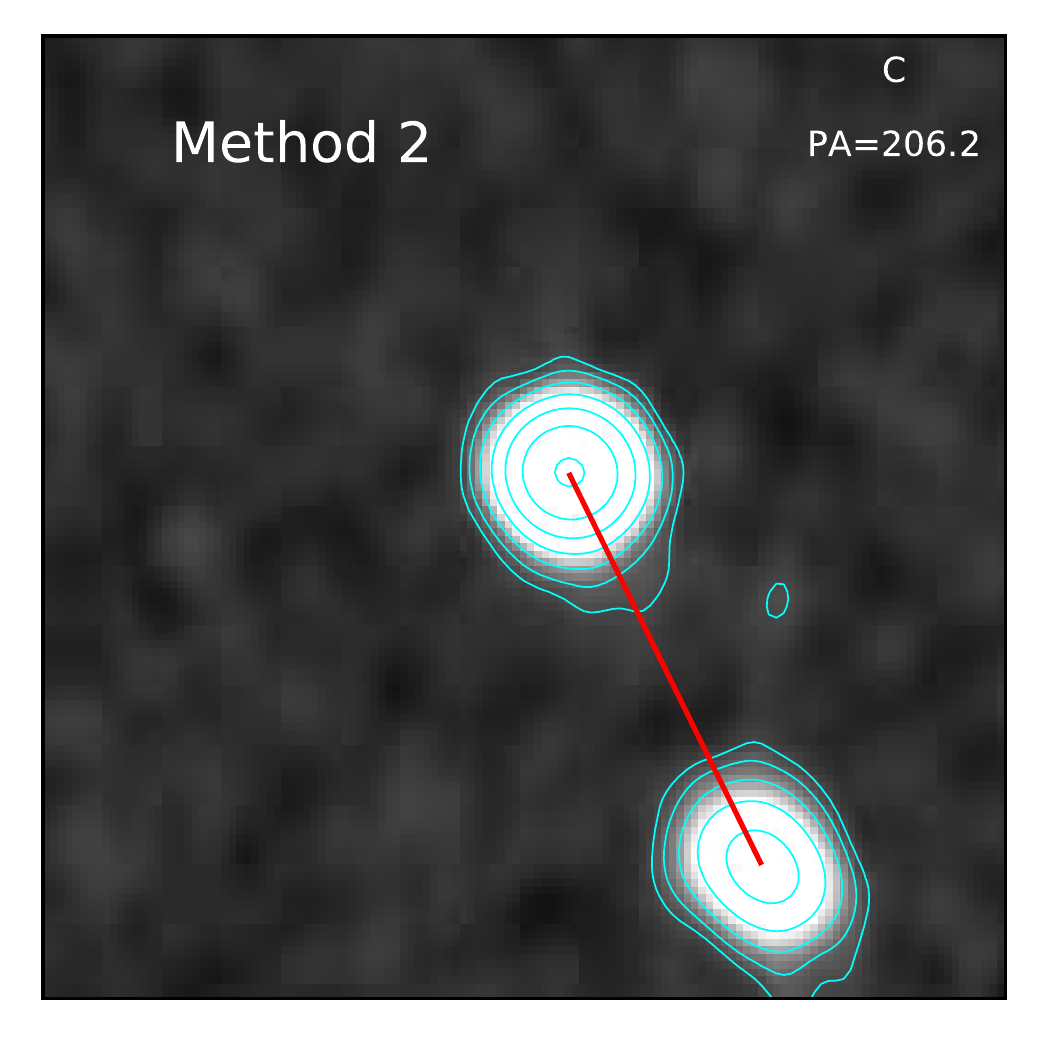}&
   \includegraphics[trim={0.0cm 0.0cm 0.3cm 0.0cm},clip,width=0.31\textwidth]{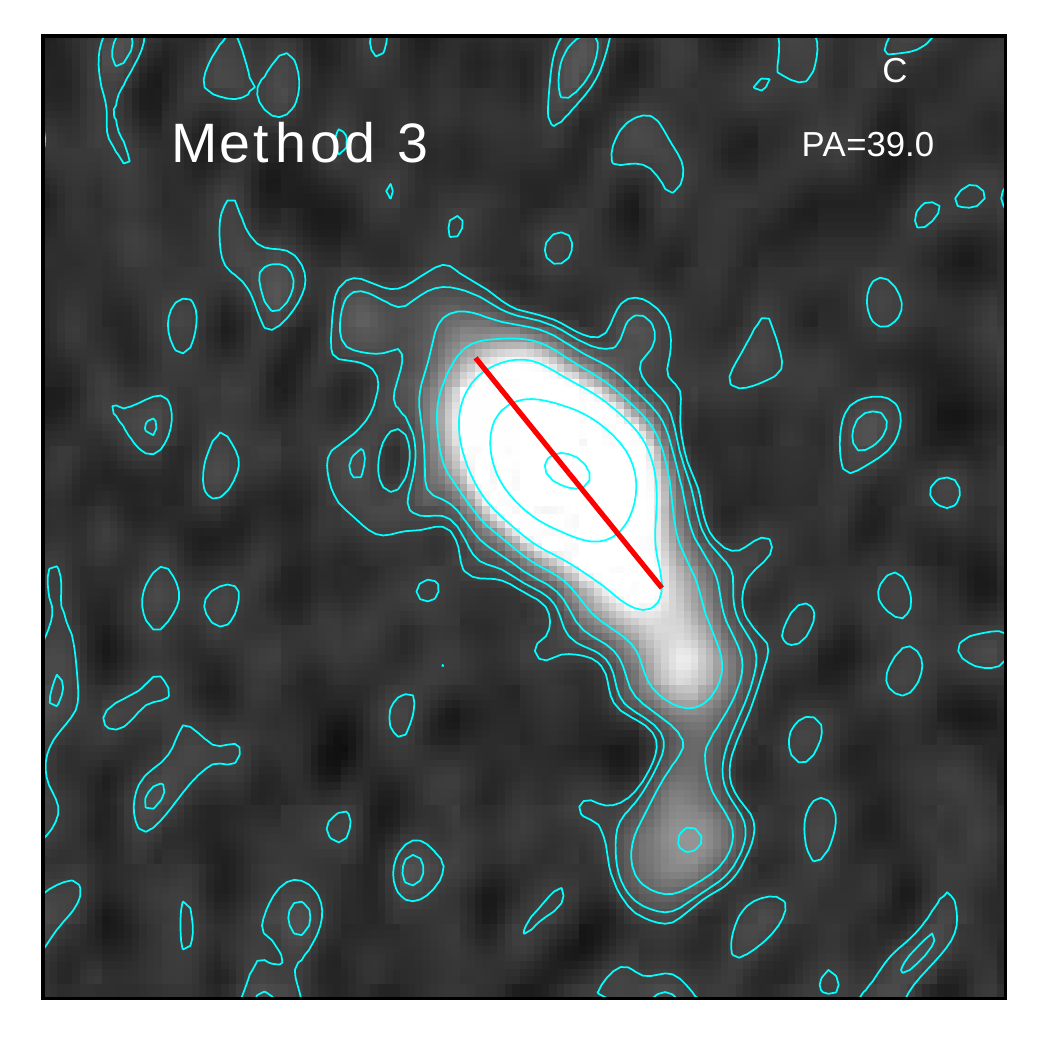}\\
   \includegraphics[width=0.31\textwidth]{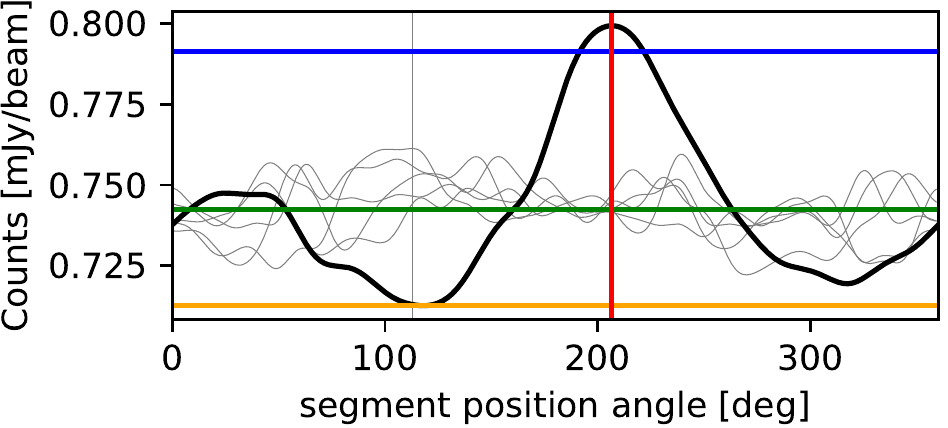}&
   \includegraphics[width=0.31\textwidth]{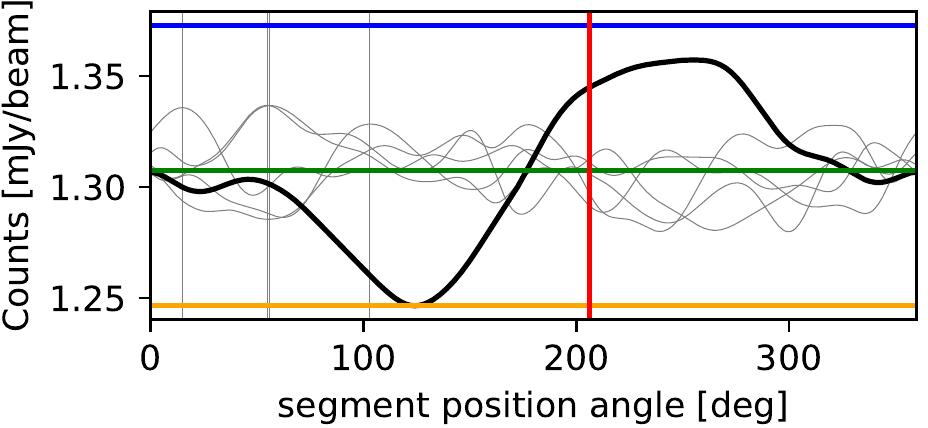}&
   \includegraphics[width=0.31\textwidth]{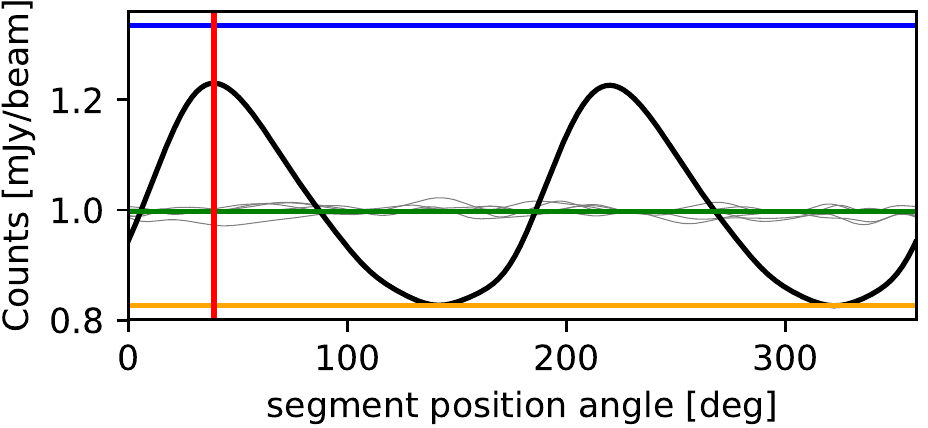}\\
    \end{tabular}
    \caption{Examples of the jet detection with the 3 methods. The left column shows the C-band image of J0933+6106 
(RA=09h33m10.4s, DEC=+61d06m46s) (top) and its azimuthal flux density distribution within the 
$3.5\,\sigma$ circular area (bottom). The central column presents similar plots for J1005+2403 
(RA=10h05m07.9s, DEC=+24d03m38s). The right column presents similar plots for J1524+2900
(RA=15h24m05.2s, DEC=+29d00m23s). The red lines indicate the determined PA$_{\rm jet}$. The solid 
black line shows the azimuthal brightness distribution for the source, and five gray lines show the 
same value calculated in random positions within 40 mas from the source. All three images are from the 
VLBA Imaging and Polarimetry Survey \citep[VIPS,][]{Helmboldt2007}. The levels in all images correspond to 1.5, 3, 6, 12, 24, 48 and 96 percent of the maximum.}
    \label{fig:alg}
\end{figure*}

In the processed images, we fit the central source using a circular Gaussian, with a standard deviation $\sigma$. We defined a stripe starting at the center and pointing to the north pole of the frame, $3.5\,\sigma$ in length and 2 pixels in width. We measured the azimuthal distribution of surface brightness by rotating the image 720 times with interpolation in steps of 0.5$\dg$ and summing the counts within the stripe. We repeated this process for all sources and at all frequencies. Some examples of such distributions are presented in the bottom panels of Fig.~\ref{fig:alg}. Due to the variety of morphological features that can be distinguished in images of radio jets, we employed different approaches in order to successfully identify the jet directions. As a first method, a jet was considered detected when the maximum of the flux density distribution exceeded 2.4 times its standard deviation (left panels of Fig.~\ref{fig:alg}). In this case, the determined PA of the jet was the position angle which corresponded to the maximum of the distribution. If the $2.4\,\sigma$ limit was not reached, a second method was used. Firstly, the circular area extending $3.5\,\sigma$ away from the center was masked and the flux density peak of the remaining area was determined. In the case when the peak value was above $3\,\sigma$ compared to the local background, another circular Gaussian was fit, centered at the position of the maximum. If the fitting procedure converged and the Gaussian width was within the range 0.6\;mas < $\sigma$ < 5\; mas, then the jet PA was determined as the position angle of the Gaussian center. This method is particularly useful in the cases where the image shows a symmetric core at the center and an isolated knot farther out, such as the one shown in the middle panels of Fig.~\ref{fig:alg}. During the visual inspection of images of sources studied in our previous paper \citep{Blinov2020} we identified 484 sources where the jet was not detected by either of the two methods described, but was visibly present in the images. In order to handle a major part of these sources we used a third method, which is able to find the jet position angle of sources with two peaks in the opposite directions of the azimuthal distribution of brightness. Such distribution is characteristic for two-sided jets. However, it can also be found in some cases where the counter jet is not prominent. An example of such source is shown in the right panels of Fig.~\ref{fig:alg}. In such a case, the $2.4\,\sigma$ limit of the first jet-detection method can not formally be reached due to a double symmetric peak in the azimuthal brightness distribution. Therefore, if neither of the first two methods identified a jet, the image was rotated 360 times with interpolation in steps of 0.5$\dg$ and for each of these rotations the counts were summed in a stripe that had twice the length of the stripe used in method 1, and was centered at a source peak. Additionally, a few tens of sources with obvious prominent jets that were not detected by either of the three methods were reprocessed after a slight tuning of the free parameters, which resulted in visually satisfactory jet detection. 

Outputs of the automatic jet determination methods were visually examined by the authors. We set up a web service conceptually resembling interfaces of citizen-science projects \citep[e.g.][]{Banfield2015}. At each web-page reload, a user was shown the results for a random source, as the ones presented in Fig.\ref{fig:alg}. Plots and images at all available bands were given simultaneously. The images were displayed in JS9\footnote{\url{https://js9.si.edu/}} windows, which allow the user to adjust several parameters such as the dynamic range levels, the zoom, the scale etc. Users could judge the results of the jet detection algorithm for each band using four buttons positioned underneath it, "good", "bad", "no jet" and "unclear". Every source was inspected at least by one of the authors. The total number of images inspected was 29821. In 10397 cases the authors confirmed the jet PA obtained through one of the methods (labeled "good"); in 2701 sources the authors did not confirm the determined jet direction ("bad"); 14799 were the sources without a visible jet ("no jet") and in 1924 it was not clear whether there was a real jet or some artifact resembling a jet ("unclear"). In a conservative approach, we used only "good" sources for the further analysis. The number of sources for which we have jet PAs is 7290.

\section{Local alignments of jets position angles}\label{sec:align}
\subsection{Parallel transport}\label{subsec:partrans}
\begin{figure}
   \centering
   \includegraphics[width=0.48\textwidth]{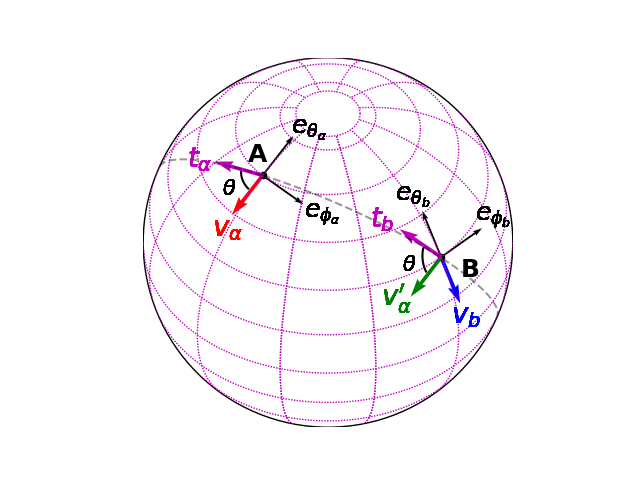}
      \caption{Demonstration of the parallel transport method. Sources lying at the points A and B both have their vectors $\Vec{v_a}$ (red) and $\Vec{v_b}$ (blue) point to the South Pole and thus have position angles $\xi_a=\xi_b=180\dg$. However, they are not aligned. In order to compare them $\Vec{v_a}$ should be parallel-transported to location B following the great circle passing through A and B (dashed line). The resulting vector is $\Vec{v_a'}$ (green). $\Vec{t_a}$ and $\Vec{t_b}$ are the vectors tangent to the sphere and parallel to the plane of the great circle at locations A and B respectively. The angle ($\Vec{v_a}$, $\Vec{t_a}$) is equal to the angle ($\Vec{v_a'}$, $\Vec{t_b}$) in order to ensure that $\Vec{v_a}$ is parallel to $\Vec{v_a'}$. $\Vec{e_{\theta}}$ and $\Vec{e_{\phi}}$ are the unit vectors pointing northward and eastward. The two sources have different reference systems.}
         \label{fig:partrans}
\end{figure}

Jet directions can be described as unit vectors on a sphere (since we examine only projections of the jets, we consider them to be tangent to a sphere), with their origin at the position of their respective source. They can be fully characterized by the jet position angle, which is measured with respect to the local meridian. For this reason, two sources lying on the sphere in different locations cannot be directly compared, as their reference system is not the same. For example, the PA vectors of the two sources shown in Fig.\ref{fig:partrans} (red, blue) both point to the South Pole (PA = $180\dg$) in their local reference frame, however they are not aligned. In order to compare vectors lying in different positions on a sphere, we have to use the parallel transport method.

Parallel transport is the process of moving a vector (or any geometrical data) along a path of a connected manifold, keeping it constant to the local geometry. While transporting a vector in Euclidean space is trivial, the result of the parallel transport between two points on a sphere will depend on the path taken \citep[e.g.][]{carroll_2019}. In order to be able to compare two vectors in curved space they need to be in the same tangent space. In our case, that means to parallel transport one vector to the position of the other following the great circle that passes through both points. This way, we can calculate a coordinate-invariant inner product between them.

We therefore consider the great circle connecting two sources, say $A$ and $B$, whose PA vectors and PA values are denoted as $\Vec{v_a}$, $\Vec{v_b}$ and $\xi_a$, $\xi_b$ respectively (see Fig.~\ref{fig:partrans} for visualization). We need to transport one vector to the location of the other, so let $\Vec{v_a}$ be transported to the position $B$. We determine the vector $\Vec{t_a}$ in the position $A$ which is tangent to the sphere and parallel to the plane of the great circle involved in the operation. Similarly, we can determine $\Vec{t_b}$ for point $B$. Translation of $\Vec{v_a}$ to $B$ should happen in such a way that the vector $\Vec{v'_a}$ resulting after parallel-transport is parallel to the initial vector $\Vec{v_a}$. Therefore, the angle ($\Vec{v_a}$, $\Vec{t_a}$) should be equal to the angle ($\Vec{v'_a}$, $\Vec{t_b}$). We parametrize the sphere using spherical coordinates ($r$,\,$\theta$,\,$\phi$), with the corresponding unit vectors ($\Vec{e_r}$,\,$\Vec{e_{\theta}}$,\,$\Vec{e_{\phi}}$), where $\Vec{e_r}$ points to the center of the sphere, $\Vec{e_{\theta}}$ points northward and $\Vec{e_{\phi}}$ eastward. Then the new position angle $\xi'_a$ of the source $A$ will be
\begin{equation}
    \xi'_a = \xi_a + \chi_b - \chi_a,
\end{equation}
where $\chi_a$ is the angle between $\Vec{t_a}$ and $\Vec{e_{\theta_a}}$ and $\chi_b$ the angle between $\Vec{t_b}$ and $\Vec{e_{\theta_b}}$. For a detailed description of the mathematical operations taking place in the process we refer readers to \cite{Jain2004} and \cite{Contigiani2017}.

\subsection{Statistical test}\label{subsec:s_test}
\label{SandZ_tests}

In order to evaluate potential alignments of jet directions in certain areas of the sky, we employed an adjusted version of the S-test, which has been widely used for similar studies.
The S-test was firstly introduced by \cite{Hutsemekers1998} to statistically characterize alignments of polarization vectors on a spherical surface and can be used for similar quantities such as jet position angles. This test relies on measuring the dispersion of PAs for a group of neighbouring sources among the main sample. Later, \cite{Jain2004} developed an equivalent but computationally-faster version of the test, which we used in this work. While in said studies the interest was focused on global effects, we probed alignments in local regions, each time focusing in a subgroup of the main sample. Therefore, we employed a slightly modified version of the \cite{Jain2004} algorithm as described below. We use the same notation as in~\ref{subsec:partrans}.

Given a region of interest $i$ which encompasses $n$ sources, we look for a position angle $\xi$ lying in the center of the region which best describes the average position angle of the sources. We therefore consider their PA unit vectors $\Vec{\xi_1}, \Vec{\xi_2}, ..., \Vec{\xi_n}$ to define the following quantity
\begin{equation}\label{eq. dispersion}
    d_{i,n}(\xi) = \frac{1}{n}\sum_{k=1}^n(\Vec{\xi},{\Vec{\xi_k'}})
\end{equation}
where $\Vec{\xi'_k}$ denotes the new PA unit vector of the $k_{th}$ source, parallel-transported to the center of the region and $\Vec{\xi}$ the unit vector of $\xi$. $d_{i,n}(\xi)$ represents the level of alignment of the sources with $\xi$.
The angle that depicts the mean direction of the jet PAs is the one that also maximizes this expression, and, therefore, the significance level of alignment.
In this case, the function of Eq.~\ref{eq. dispersion} takes the value
\begin{equation}
D_{i,n} \equiv d_{i,n}\biggr\rvert_{max} = \frac{1}{n} \left[ \left(\sum_{k=1}^n \cos{2 \xi'_k} \right)^2 + \left(\sum_{k=1}^n \sin{2 \xi'_k} \right)^2 \right]^{1/2},
\end{equation}
where larger values of $D_{i,n}$ correspond to better alignment. The extreme cases are $D_{i,n}=1$ which implies perfect alignment and $D_{i,n}=0$ which corresponds to complete uniformity. We define the significance level (S.L.) of the alignment similar to \cite{Contigiani2017}:
\begin{equation}
    S.L.=1-\Phi\left(\frac{D_{i,n}-\langle D_{i,n} \rangle_{MC}}{(\sigma_{i,n})_{MC}}\right),
    \label{eq:SL_S}
\end{equation}
where $\Phi$ is the normal cumulative distribution function. $\langle D_{i,n} \rangle_{MC}$ is the average value of $D_{i,n}$ for the $i_{th}$ region, found through Monte-Carlo (MC) simulations, assuming random distribution of PAs in the sky. Similarly, $(\sigma_{i,n})_{MC}$ is the standard deviation of the simulated $D_{i,n}$ for the same region. MC simulations were carried out as follows. We assigned new PA values to all sources of our sample, drawn from the uniform distribution. Subsequently, we measured $D_{i,n}$ for the region, using the random PA values assigned to the sources within it. From 10000 repetitions we derived the average and standard deviation, which we used for the computation of S.L..

$3\,\sigma$, $4\,\sigma$ and $5\,\sigma$ confidence levels correspond approximately to values of $log(S.L.)=-2.57$,  $-4.20$ and $-6.24$ respectively, while $log(S.L.)<-6.24$ corresponds to more than $5\,\sigma$ confidence.

\subsection{Search for alignments}\label{subsec:alignments} 
The search for alignments was performed with two different ways of binning the parameter space (sky coordinates and redshift) in order to probe regions with different characteristics.

\begin{figure}
   \centering
   \includegraphics[width=0.48\textwidth]{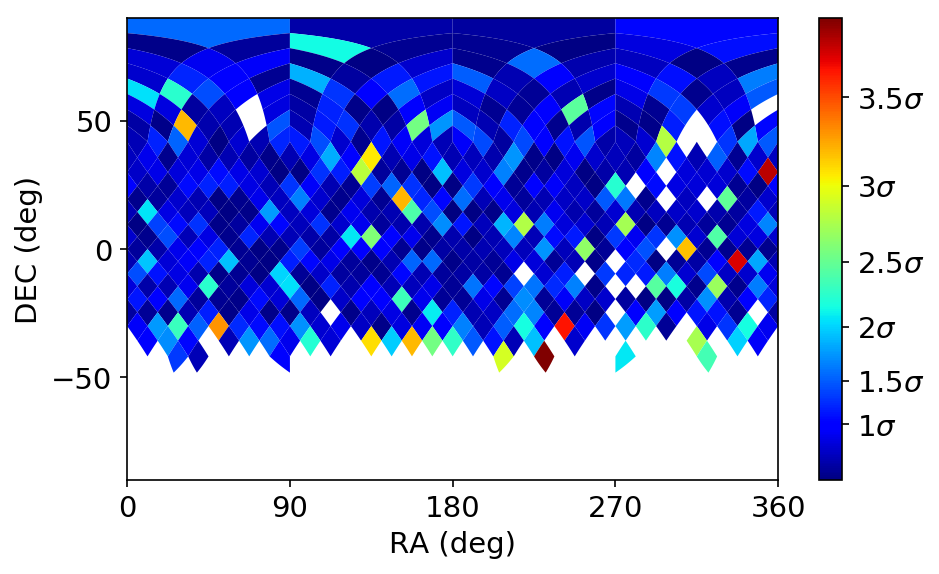}
      \caption{HEALPix map with NSIDE=8 indicating with color the level of alignments significance in each pixel for sources with any redshift. Significance levels are derived based on Eq.~\ref{eq:SL_S}}
         \label{fig:healpix}
\end{figure}

The first method is based on HEALPix tessellations of the angular coordinates \citep{Gorski2005}. HEALPix creates discretized maps to allow for fast scientific computations by dividing the sphere in equal-area pixels. The sphere is partitioned in a number of base-level tessellations, each of which can then be divided in $n\times n$ sub-elements, thus creating a hierarchical structure. Numerous spherical tessellations can be constructed, depending on two parameters: $N_{\theta}$, which is the number of base-resolution pixel layers between the poles of the sphere, and $N_{\phi}$, the number of circumpolar base-resolution pixel layers. Therefore, the total number of base-resolution pixels is $N_{base-pix}=N_{\theta} N_{\phi}$, each encompassing an area of $A_{base-pix}=4\pi/N_{base-pix}$ steradians. The standard definition of the HEALPix full-sky map uses $N_{\theta} = 3$ and $N_{\phi} = 4$. The resolution of the grid is defined by the number of partitions along the side of a base-level pixel, $N_{side}$. It follows that the total number of pixels in a HEALPix map is $N_{pix}=12N_{side}^2$, each covering an area of $A_{pix}=\pi/(3N_{side}^2)$ steradians.
We performed the search for alignments in each pixel of the discretized map, for different redshift cuts, using 5 different values for $N_{side}$: 2, 4, 8, 16, 32 which roughly correspond to areas encompassed by circles of radius 16.5, 8.3, 4.1, 2.1, 1.0 degrees respectively. For each value of $N_{side}$ we measured the S.L. of alignment of the sources that fall within each pixel as described in Sect.~\ref{subsec:s_test}, if the pixel contained at least 5 sources. An example of measured S.L. in a HEALPix map with $N_{side}=8$ and no redshift cut is shown in Fig.~\ref{fig:healpix}. We repeated the procedure for all possible redshift cuts in the range [0, 5] with step 0.1.

The second method is similar in concept with the one described above. We divided the virtual sphere in meridians and parallels, with the angular distance between sequential meridians in the equator being equal with the angular distance between sequential parallels. In each interception of a parallel with a meridian, we placed a circle with radius equal to the angular distance to the next parallel. This allowed neighbouring regions to partially overlap, which would not be possible with the HEALPix approach, and ensured that we did not miss regions of high alignments significance. We repeated the process for all possible redshift cuts in the range [0, 5] with step 0.1 and setting different angular distances between sequential parallels to the following values 5, 6, 7, 11 and 13 deg. We used these values to cover angular scales which cannot be probed with the HEALPix representation. Within every circular region containing at least 5 sources, we studied the distribution of PAs and measured the significance of possible jet alignments. When the alignments were detected with a significance of more than $4\,\sigma$ the regions were kept for further analysis. The search with these two methods returned eight regions.

Furthermore, we performed the search for alignments once again with both methods, but without imposing redshift cuts. With this approach, we detected 64 regions with more than $4\,\sigma$ significance of alignment. However, the positions of all of these regions are in low declination, where we expect our data to suffer from instrumental biases (see discussion in Sect.~\ref{subsec:bias}) and thus we discard them without further consideration.

After the initial selection, and for each of the retained regions, we searched for the optimal region definition such that the significance of the jet-PA alignment was maximized. During the optimization, performed with the pyswarms\footnote{\url{https://pyswarms.readthedocs.io/en/latest/intro.html}} Python implementation of the Particle Swarm Optimization algorithm, both the radius and the center of the circular regions were allowed to vary together with the lower and upper redshift limits. The radius was allowed to vary in the range from 0.5 to 2 times its initial value, the center of the region could be shifted in any position within the initial circular region and the redshift limits were allowed to vary freely in entire range from 0 to 5.

\section{Results and discussion}\label{sec:results}
\begin{figure*}
   \centering
   \includegraphics[width=0.9\textwidth]{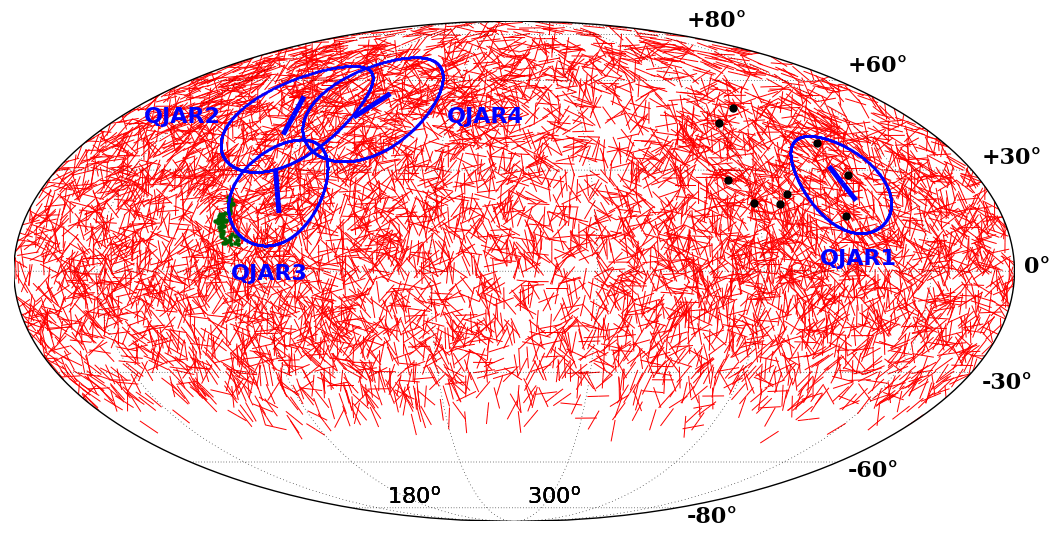}
      \caption{Positions of the strong-alignment regions QJARs ($S.L.>5\,\sigma$) along with the average direction of PAs within each region are plotted with blue. Red segments represent the PA of each source. Green points represent the positions of the members of the Huge-LQG \citep{Clowes2013}. Black points represent the positions of the members of the Giant GRB Ring \citep{Balazs2015}.}
         \label{fig:regions_galactic}
\end{figure*}

After optimization of the initial regions, six our of eight showed an alignment with $> 5\,\sigma$ significance. For future reference we shall call these regions QJAR (quasars jet alignment region). We decided to discard regions with $S.L.< 5\,\sigma$ in order to ensure that the the probability $\wp$ that they arise by chance is extremely low ($\wp\lesssim 6\times10^{-7}$). In what remains we explore the possibility that the observed alignments reflect systematic effects in our data set. As we shall see in Sect.~\ref{subsec:bias}, we cannot fully discard this possibility for two out of six QJARs. We discuss in more details the four remaining regions for which systematic effects are unlikely to be responsible for the very high significance of the alignments and explore a possible connection with other known large-scale anomalies in Sect.~\ref{subsec:cosmic_struc}. We present  the region characteristics in Table~\ref{tab:reg} and their sizes and locations on the sky in Fig.~\ref{fig:regions_galactic}.

\begin{table*}
\caption{ Regions with significant alignments. Col. 1: Region identifier Col. 2,3: Coordinates of the center of the region. Col. 4: Radius of the region. Col. 5,6: The lower and upper bounds in redshift of the sources in the region. Col. 7: Significance Level in $\sigma$. Col. 8: Average PA of the sources, after parallel-transporting them to the center of the region.}             
\label{tab:reg}      
\centering                          
\begin{tabular}{c c c c c c c c c}        
\hline\hline                 
ID & RA & DEC & Radius & $z_{\rm min}$ & $z_{\rm max}$ & S.L. &  $PA_{avg}$\\    
   & deg & deg & deg & & & $\sigma$ & deg\\    
\hline                        
QJAR1& 36.18&25.95& 14.82&0.02&1.53&5.54&$-$16.9  \\
QJAR2& 168.75&47.54&18.11&0.36&1.89&6.74&$-$17.7  \\
QJAR3&180.15&23.56&16.07&0.29&0.65&5.72&$-$25.6  \\
QJAR4&201.83&50.97&18.27&0.07&0.35&5.38&37.6  \\

\hline                                   
\end{tabular}
\end{table*}

\subsection{Possible systematic bias}\label{subsec:bias}

\begin{figure*}
   \centering
   \includegraphics[width=0.9\textwidth]{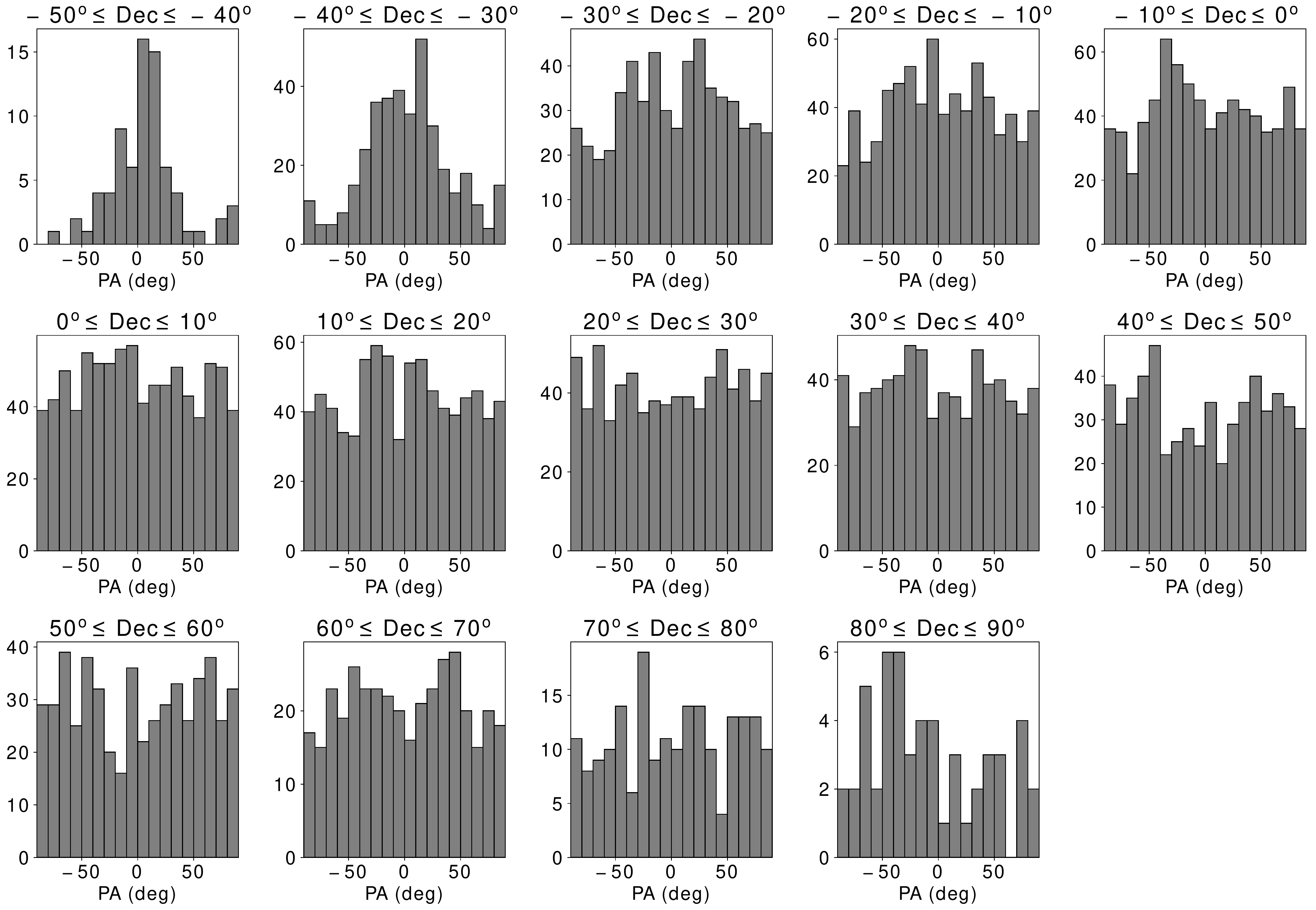}
      \caption{Distributions of jet PAs in different declination slices.}
         \label{fig:all_hists}
\end{figure*}

Statistical alignments of PAs could potentially arise due to instrumental effects. Here we explore a known systematic that could affect our results.The configuration of the VLBI array can introduce distortions in the observed images. Due to the specific location of the major VLBI telescopes, radio images are typically better resolved in the East-West than in the North-South direction. This makes the restoring beam elliptical, and in some cases very elongated along the North-South direction, and may lead to a source structure with a hint of extension North-South, purely produced by the restoring beam. Although this was taken into account during the data collection and reduction process as described in Sect.~\ref{sec:data}, there may still remain a bias owning to this effect. Moreover, since the majority of the VLBI antennas are located in the Northern hemisphere, images of the South-most targets are more prone to exhibiting elongated structures as a result of instrumental bias. This possible systematic effect is presumably redshift independent since the VLBI beam shape is agnostic to the source distances. In fact, such an effect would be most prominent when considering all the sources of a region, without imposing any redshift restrictions. In order to investigate the role of this potential bias in our data, we made slices in declination and examined the distribution of jet directions within each slice. We present them in Fig.~\ref{fig:all_hists}. Noticeably, sources with DEC $\lesssim -30 \dg$ show radio structures preferentially aligned in the North-to-South orientation ($\rm PA=0\dg$), as expected from the described systematic effect. This indicates that, despite the treatment applied in Sect.~\ref{sec:data} to correct for non-circular beam effects, systematics are still likely to be present at the PA level. For this reason we rejected the 64 regions found without redshift cut as described in Sect.~\ref{subsec:alignments}.
However, this trend seems to wear off for sources located further North.

Out of the initial six QJARs, two were centered at DEC $\lesssim -20\dg$, encompassing sources well below DEC $\lesssim -30\dg$. Following Fig.~\ref{fig:all_hists} and the arguments above, we decided to discard these regions, as we cannot be certain that the alignment signal is not affected by instrumental effects. For the remaining four regions, we emphasize that the alignments are spatially delimited in redshift, and sources outside the redshift range, with similar RA and DEC, do not show the same trend in jet directions. This indicates that i) the alignments occur in specific areas of the 3D space and ii) instrumental effects are not likely to be responsible for the observed trends.

In order to further test the robustness of the remaining QJARs against instrumental bias, we compared the distribution of jet PAs within each region with the distribution of beam orientations ($PA_{beam}$) for the same sources. In the case of multiple observations in a particular source the $PA_{beam}$ is just the average value between different epochs, though, for the same source, $PA_{beam}$ is almost identical in all epochs. In Appendix~\ref{appendix} we present the comparison of $PA_{jet}$ with $PA_{beam}$ for the jets within each region together with a sky plot and the redshift distribution of the sources. Wee found that $PA_{jet}$ do not follow $PA_{beam}$ for either of the four QJARs. Moreover, they are located in the Northern hemisphere and, following the distributions presented in Fig.~\ref{fig:all_hists}, should be free of instrumental contamination. Therefore, we argue that, for these regions, the effect of alignment is most probably natural and not due to instrumental biases or artifacts.

\subsection{Comparison with known cosmic structures}\label{subsec:cosmic_struc}
We explored any potential correlation between our QJARs and other known structures or cosmic anomalies. We examined whether our regions are co-spatial with the "hot" or "cold" areas of the CMB dipole, quadrapole and octupole \citep{Rassat2014}. We cross-matched our regions with positions of Sunyaev-Zel'dovich galaxy clusters \citep{Planck2016} and CMB lensing density maps \citep{Planck2020} and we compared QJARs with the density map of galaxies presented by \cite{Malavasi2020}. We considered the possible quasar dipole suggested by \cite{Gibelyou2012} and explored the potential connection of QJARs' $\rm PA_{avg}$ with PAs of large quasar groups \citep{Friday2020}. In all of these cases, no obvious correlation was found.

The Giant GRB Ring \citep{Balazs2015} is considered one of the largest cosmic structures and was discovered by correlating nine GRBs located in a ring-like shape. We plot the positions of the members of the Giant GRB Ring in Fig.~\ref{fig:regions_galactic} together with the positions of the QJARs. Three of nine members are encompassed in QJAR1 and consistent with the redshift range of the sources within it. However it should be noted that the redshift span of QJAR1 ($0.02<z<1.53$) is much wider than that of the Giant GRB Ring ($0.78<z<0.86$).
We also plot in Fig.~\ref{fig:regions_galactic} the members of the Huge-LQG \citep{Clowes2013} for a size comparison. In projection, Huge-LQG is located at the border of QJAR3, however the redshift range of its members is not consistent with the sources within QJAR3, thus we cannot make any useful comparison here other than angular size. 

Finally, we cross-matched QJARs with regions of alignments found in other studies. We found that QJAR3 and QJAR4 are co-spatial and in the same redshift range as the A2 region of optical polarization vector alignments found by \cite{Hutsemekers1998}. QJAR2 partially overlaps in the coordinates-redshift space with the A1 region reported in the same study. Moreover, QJAR3 partially overlaps with the RN1 and RN2 regions of radio polarization vector alignments reported by \cite{Pelgrims2015} and QJAR2, QJAR4 also partially overlaps with the RN2 region from the same work. Notably, besides spatial correlation between QJAR4 and RN2, the average polarization angle provided for the RN2 ($\bar{\theta}=42\dg$) is remarkably similar to the average PA value for QJAR4 ($PA_{avg}=37.6\dg$).

\section{Conclusions}\label{sec:conclusions}

In this work, we investigated whether projected orientations of parsec-scale radio jets tend to align with their neighbouring sources, using VLBI images of compact objects from Astrogeo VLBI FITS image database. We developed algorithmic methods to automatically deduce directions of parsec-scale radio jets and we visually validated the products. In total, we examined 15235 sources and found obvious jets for 7290 of them. We then used the results to perform an extensive search for statistical alignments within regions of neighboring sources across the sky and found four regions (QJARs) where the significance of the effect exceeds $5\,\sigma$. We investigated in detail whether the observed alignments can be attributed to instrumental biases and we show that, most probably, the alignments are genuine. Finally, we cross-matched our results with regions of cosmological interest, such as galaxy clusters and maps of CMB moments, as well as with regions of pronounced alignments of physical quantities reported by other studies. Three out of four QJARs are found to be partly co-spatial with regions of significant alignment of either optical or radio polarization vectors. We also highlight the tantalising hints of overlap of QJAR1 with the Giant GRB Ring.

Coherence of features of neighbouring sources in three-dimensional space is often attributed to the existence of an underlying large-scale structure, which is assumed responsible for creating similarities between its members (see Sect.~\ref{sec:intro}). 
Given that most of QJARs co-exist in the coordinates-redshift space with similar regions that have been detected and confirmed by more than one studies, it is tempting to speculate about the existence of large-scale cosmic structures in these regions. If this is indeed the case with our findings, the discovered QJARs may designate some of the most sizeable cosmic structures known to date.

\begin{acknowledgements}
We thank L. Petrov who maintains the Astrogeo archive. We also thank the contributors to this rich dataset: Alessandra Bertarini and Laura Vega Garcia, Nicholas Corey, Yuzhu Cui, Leonid Gurvits, Xuan He, Yuri Y. Kovalev, Sang-Sung Lee, Rocco Lico, Elisabetta Liuzzo, Alan Marsher and Svetlana Jorstad, Christopher Marvin, Leonid Petrov, Alexandr Pushkarev, Kirill Sokolovski, An Tao, Greg Taylor, Alet de Witt, Minghui Xu, Bo Zhang and the members of the MOJAVE project team who produced the images: Dan Homan, Jose-Luis Gomez, Yuri Kovalev, Matt Lister, Alexander Pushkarev, Eduardo Ros, Tuomas Savolainen. The authors acknowledge support from the European Research Council (ERC) under the European Union Horizon 2020 research and innovation programme under the grant agreement No 771282.

\end{acknowledgements}

\bibliographystyle{aa}
\bibliography{bibliography}

%
%

\begin{appendix}  
\section{Sky plots and PA histograms of QJARs with significant alignments.}\label{appendix}

Here we present sky maps for the high-significance quasars jet alignments regions found in this work. They are complemented by distributions of the jet position angles together with the beam major axis position angles, and the distribution of sources over the redshift in each of these regions. We note that we use the original jet PA values without parallel-transporting the vectors, as the purpose of the plots is to juxtapose  $PA_{jet}$ with $PA_{beam}$, ergo parallel transport is not suitable in this case.

   \begin{figure*}
   \center
    \includegraphics[width=0.31\textwidth]{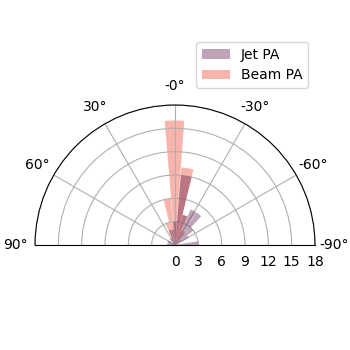}
    \includegraphics[width=0.31\textwidth]{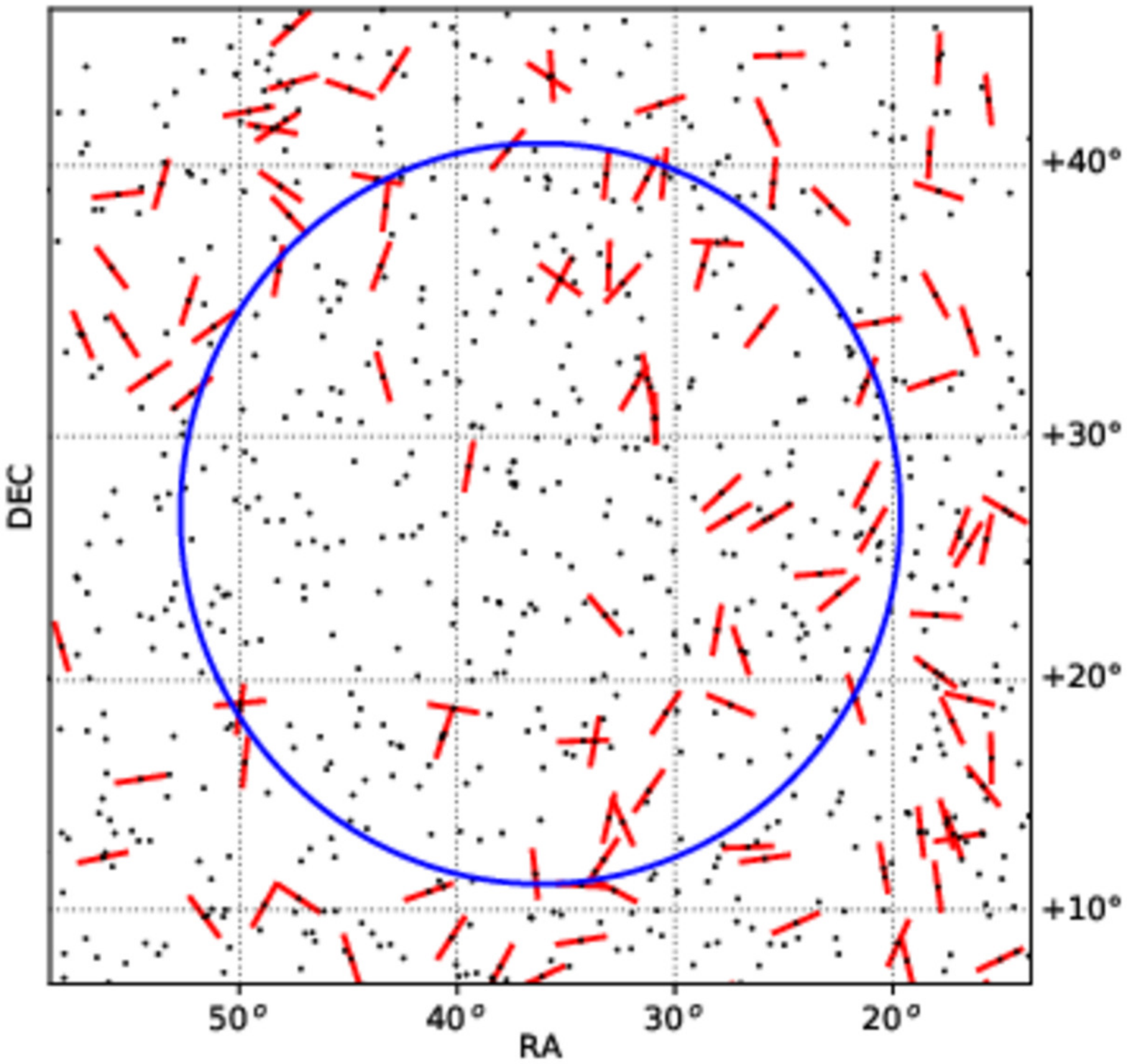}
    \includegraphics[width=0.31\textwidth]{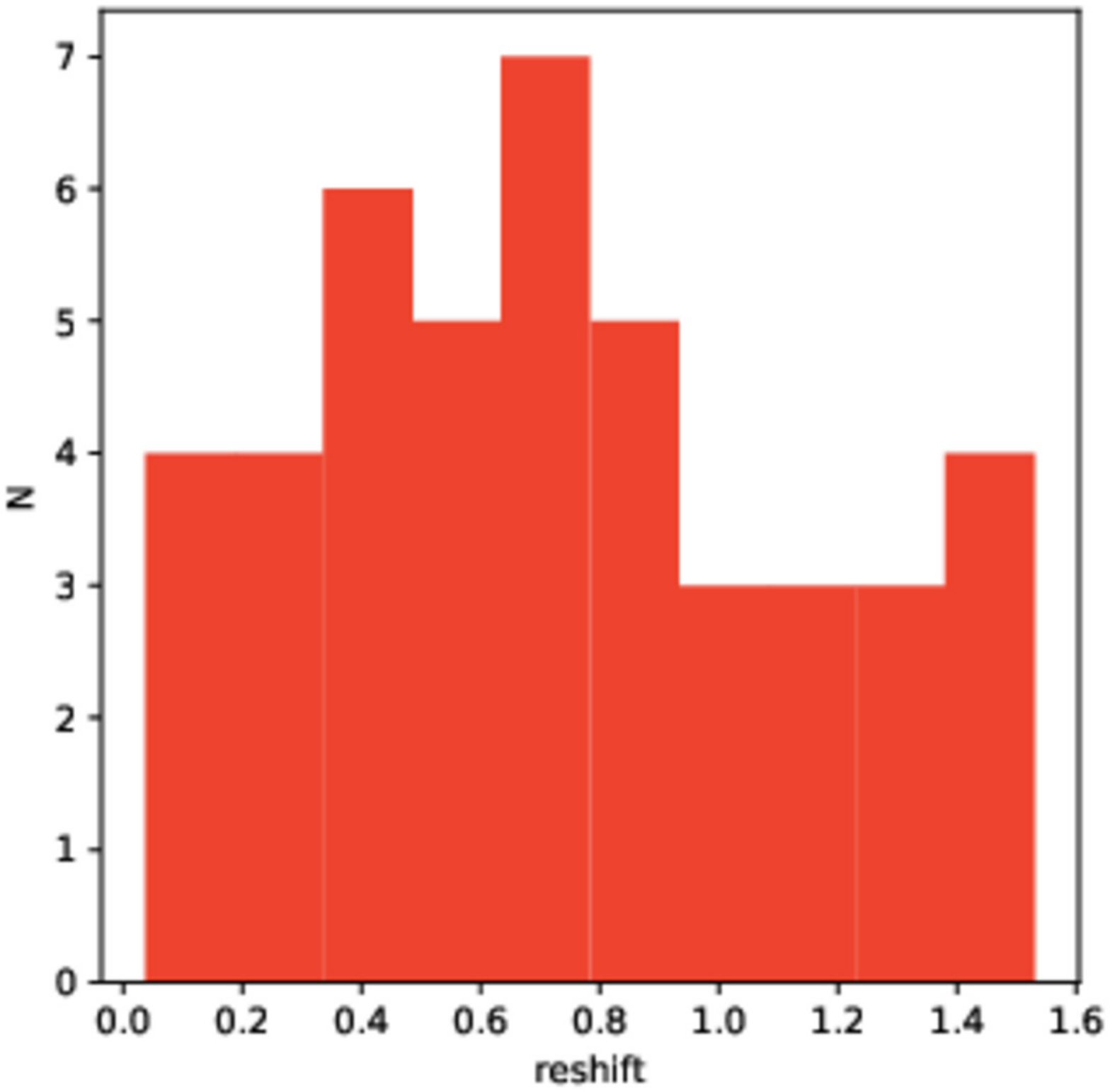}
      \caption{Distributions for QJAR1. Left panel: jet positions angles versus the radio beam orientation. Middle panel: the sky map demonstrating jets orientations (red stripes) in the region vicinity in the corresponding redshift range. Locations of sources outside the redshift range are shown by gray points. Right panel: distribution of sources over the redshift range.}
         \label{fig:reg1}
   \end{figure*}

   \begin{figure*}
   \center
    \includegraphics[width=0.31\textwidth]{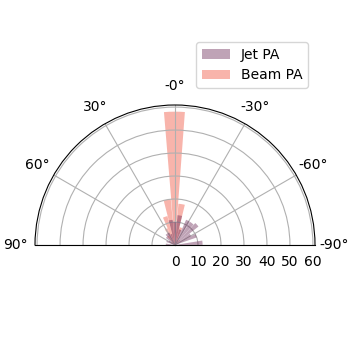}
    \includegraphics[width=0.31\textwidth]{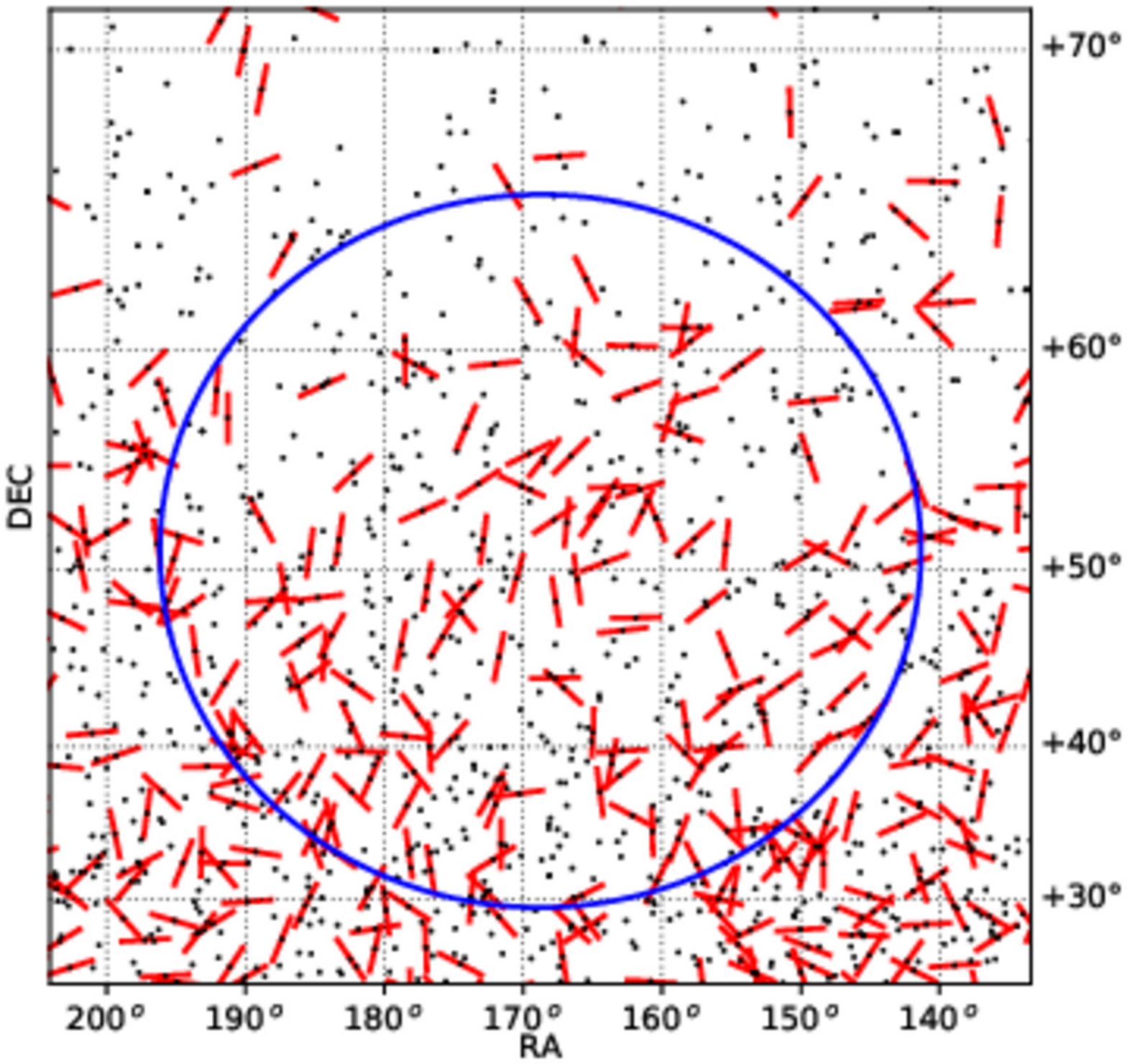}
    \includegraphics[width=0.31\textwidth]{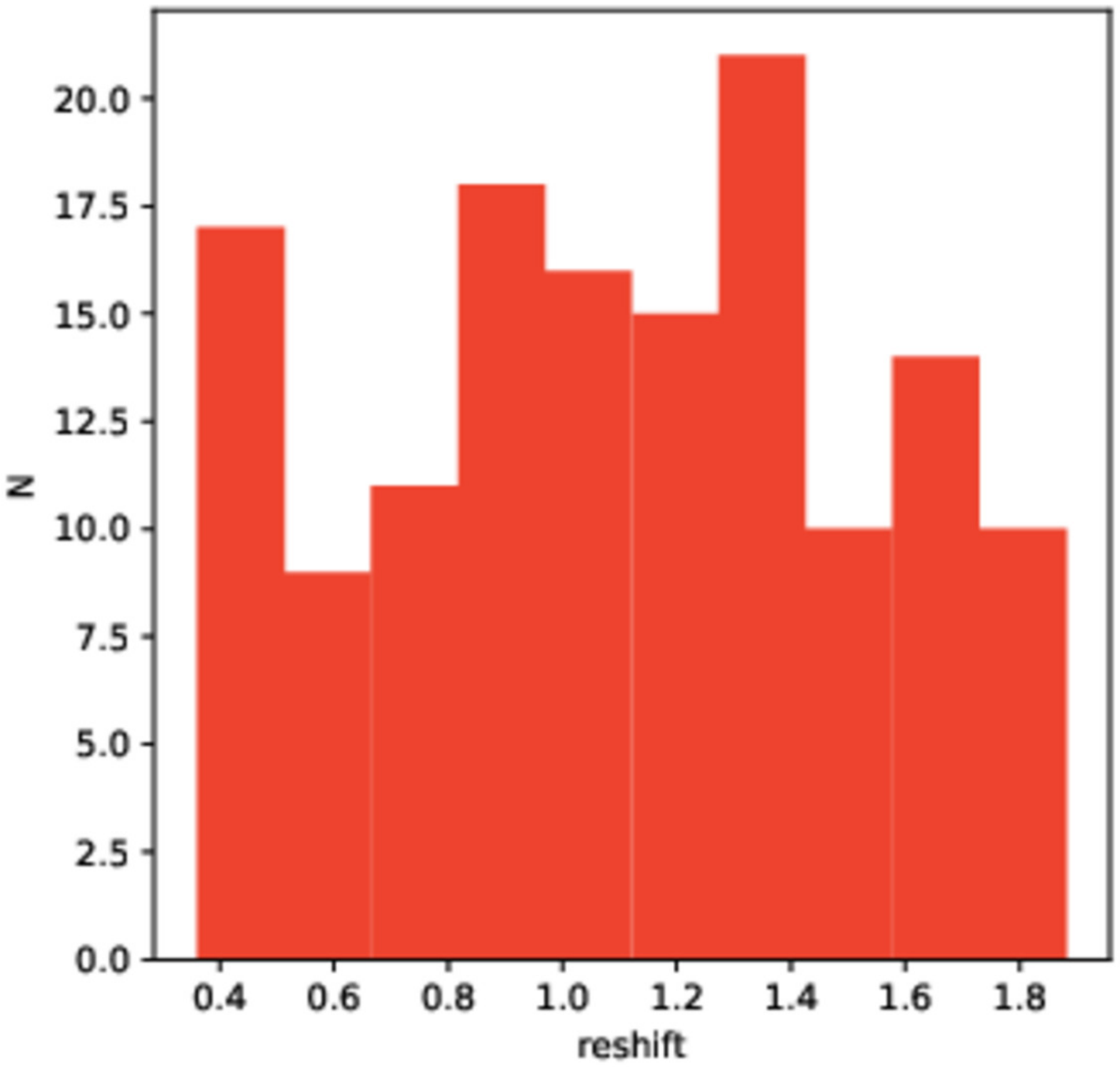}
      \caption{Same as Fig.~\ref{fig:reg1} but for QJAR2}
         \label{fig:reg3}
   \end{figure*}
   
   \begin{figure*}
   \center
    \includegraphics[width=0.31\textwidth]{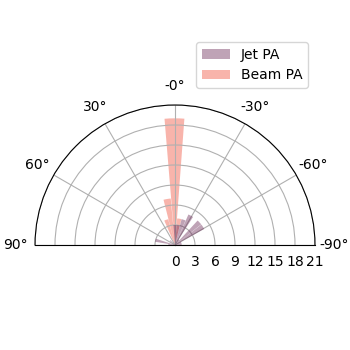}
    \includegraphics[width=0.31\textwidth]{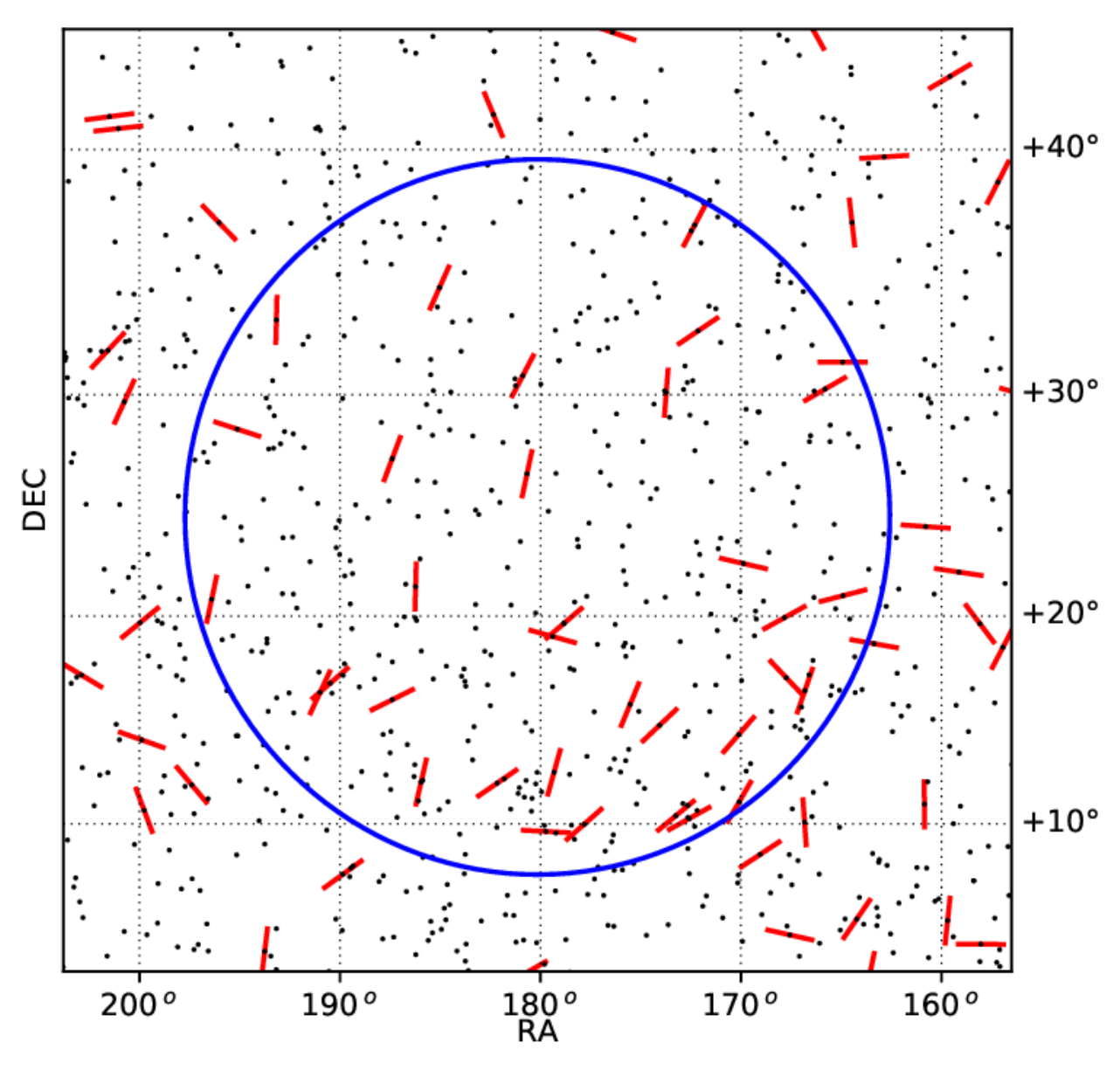}
    \includegraphics[width=0.31\textwidth]{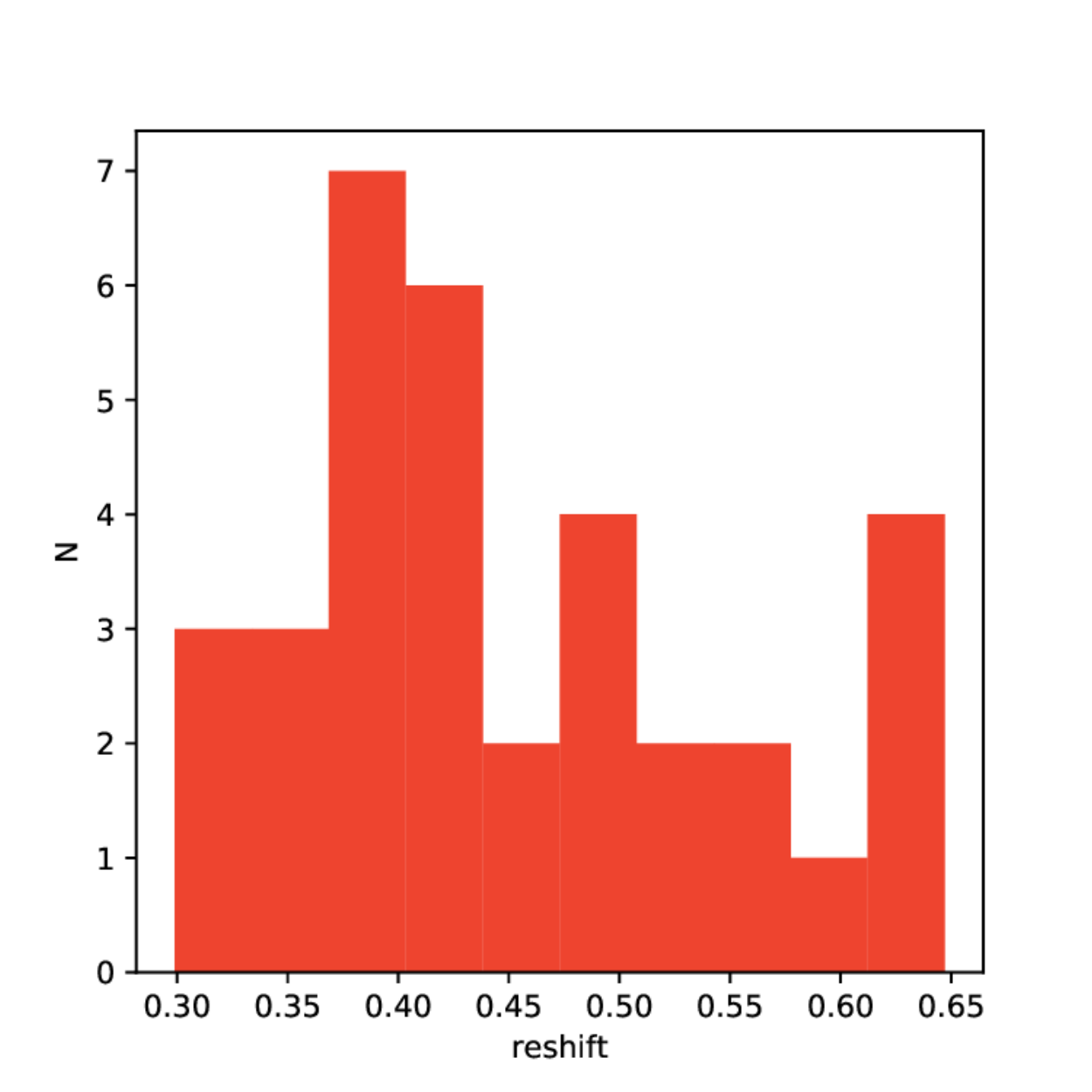}
      \caption{Same as Fig.~\ref{fig:reg1} but for QJAR3}
         \label{fig:reg4}
   \end{figure*}
   
   \begin{figure*}
   \center
    \includegraphics[width=0.31\textwidth]{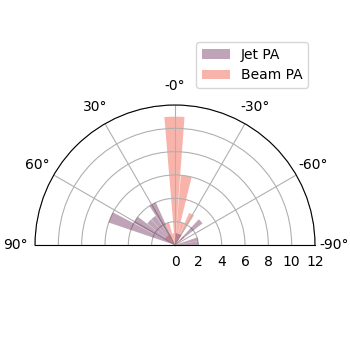}
    \includegraphics[width=0.31\textwidth]{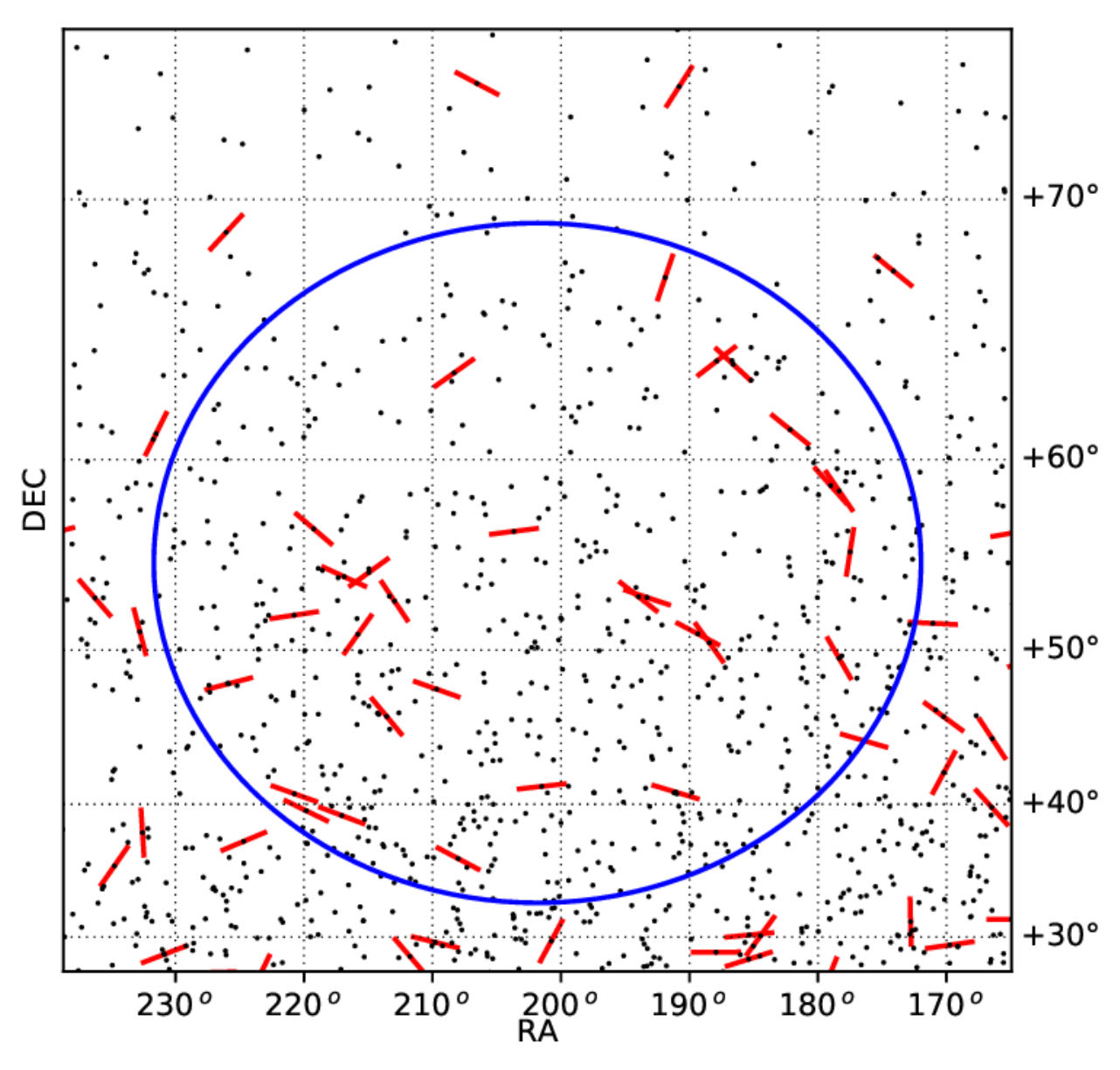}
    \includegraphics[width=0.31\textwidth]{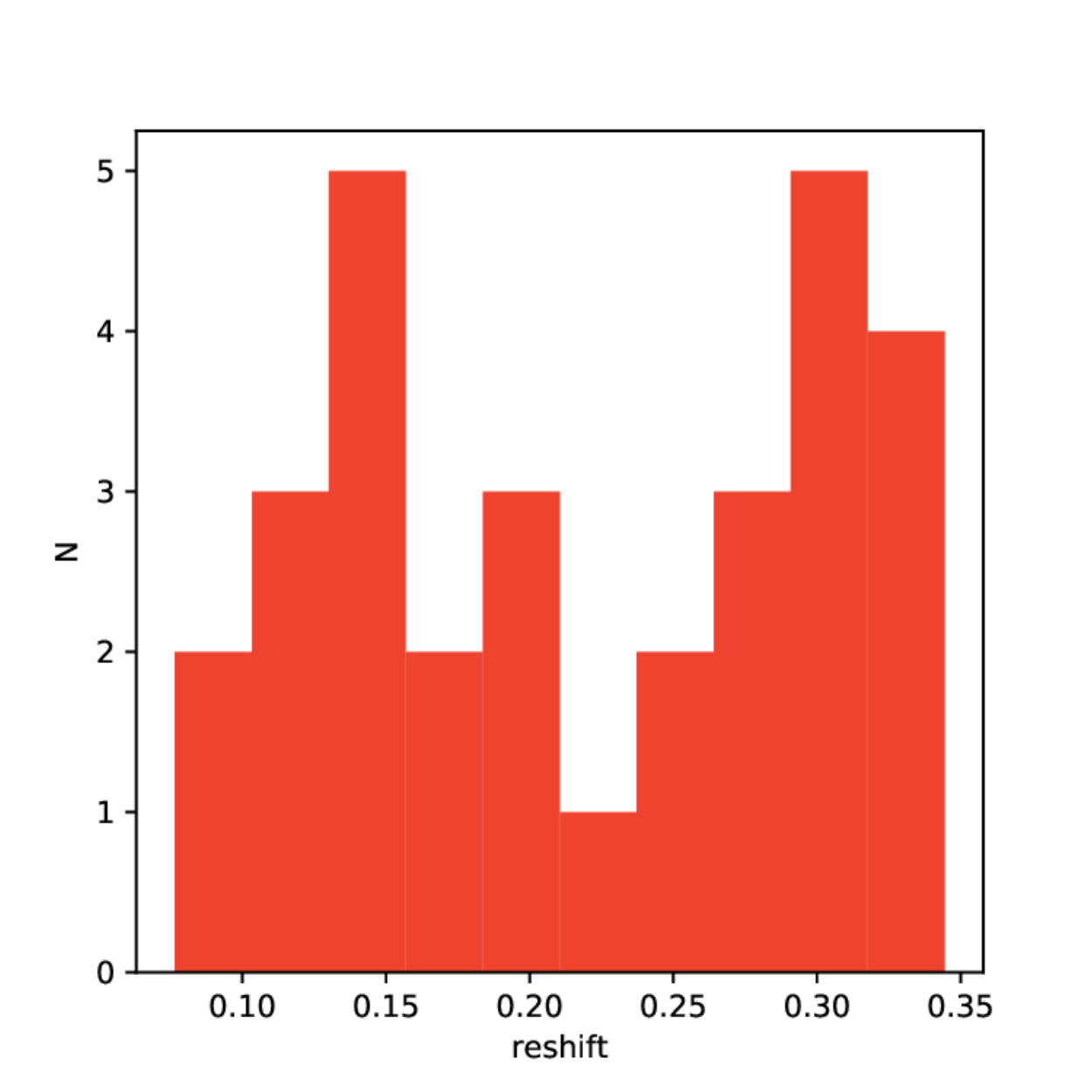}
      \caption{Same as Fig.~\ref{fig:reg1} but for QJAR4}
         \label{fig:reg5}
   \end{figure*}

\end{appendix}

\end{document}